\documentclass[pre,twocolumn,showpacs,floatfix,preprintnumbers,amsmath,amssymb]{revtex4-1}

\usepackage{amsmath}
\usepackage{amssymb}
\usepackage{graphicx}
\usepackage{dcolumn}
\usepackage{bm}

\begin{document}
\title{Active Particles Forced by an Asymmetric Dichotomous Angle
  Drive} \author{Christian Weber, Igor M. Sokolov, and
  Lutz Schimansky-Geier}
\affiliation{ Institute of Physics, Humboldt University at Berlin,
  Newtonstr. 15, D-12489 Berlin, Germany}
\begin{abstract}
  We analyze the dynamics of particles in two dimensions with constant
  speed and a stochastic switching angle dynamics defined by a
  correlated dichotomous Markov process (telegraph noise) plus
  Gaussian white noise. We study various cases of the asymptotic
  diffusional motion of the particle which is characterized by the
  effective diffusion coefficient. Expressions for this coefficient
  are derived and discussed in dependence on the correlation time and
  the intensity of the noise. The situation with a given mean
  curvature is of special interest since a non-monotonic behavior of
  the effective diffusion coefficient as function of the noise
  intensity and correlation time is found. A timescale matching condition for
  maximal diffusion is formulated.
\end{abstract}
\maketitle

\textbf{Introduction.}\;\; We study self-propelled particles moving in
two dimensions at a constant speed $v_0$. The time-dependent position
$\vec{r}(t)$ of a particle follows from integrating its velocity
$\vec{v}(t)= (v_x(t),v_y(t))=v_0\left(\cos \phi(t),\sin
  \phi(t)\right)$ with time-dependent orientation $\phi(t)$. The
orientation of the velocity vector at time $t$ is governed by
stochastic dynamics. We assume that $\phi(t)$ changes due to a constant
torque superimposed by an unbiased dichotomous
Markov process (DMP) $\zeta(t)$ which increases or decreases the local
curvature of the particle's trajectory. In addition, a Gaussian white noise is present, 
corresponding to the thermal or environmental noise of the system.

Physically, this dynamics is motivated as an approximation to recently measured bimodal
distributions $P(\Delta,\tau)$ of turning angles $\Delta=
\phi(t+\tau)-\phi(t)$ during time $\tau$, as observed in
experiments with the zooplankton \textit{Daphnia} \cite{daphniapattern,schimansky08,schimansky04}, which is also able to sustain a constant mean speed over large time scales. By using a DMP, we
approximate the bimodal structure by two delta peaks at
$(\Omega+A_-)\tau$ and $(\Omega+A_+)\tau$, where $A_-$ and $A_+$
denote the DMP strokes and $\Omega$ is the additional torque-induced angular velocity, which we will simply term ``torque'' in what follows. The constant torque can be motivated by
various biological realizations. On the one hand, there are typical
swarming characteristics which can be introduced by an effective
torque \cite{daphniapattern,pavelswarm}.  On the other hand,
an asymmetric muscularity \cite{friedrich2}, an external magnetic
field \cite{kanokov} or corresponding asymmetric boundary conditions
\cite{gautrais} can lead to an effective torque as well.

Thus, our system is fully described by the constant speed $|\vec{v}|=v_0$, and the angle dynamics  
\begin{equation}
 \dot{\phi}(t)= \Omega + \zeta(t) + \xi(t) .
 \label{curvesys}
\end{equation}
$\xi(t)$ is the Gaussian white noise with zero mean and noise intensity $D_{\xi}$.

\begin{figure}
\centering
\includegraphics[width=0.455\textwidth]{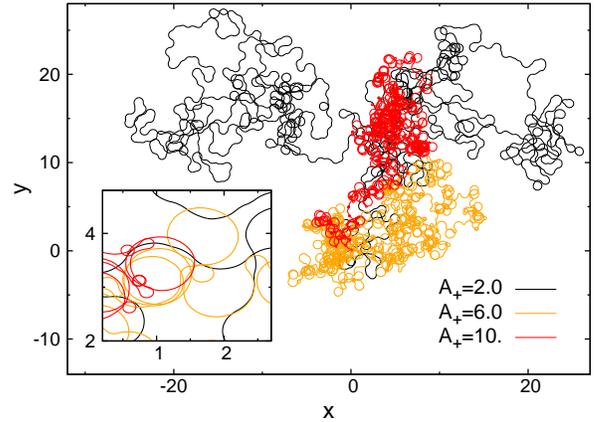}
\caption{\small (Color online) Spatial trajectories for dichotomous angular dynamics (see Eq.(\ref{curvesys}); details are shown in the inset) with $D_{\xi}=0.01$, $v_0=1$, vanishing 
torque $\Omega=0$, and the total duration $t_l=10$. Two parameters of the DMP are fixed to be $A_-=-2$, $r_+=2$, 
$A_+$ varies, and $r_-$ is defined by the vanishing mean of the DMP drive.}
\label{phiDMNtraj}
\end{figure}

The dichotomous Markov process $\zeta(t)$ is time-homogeneous and
switches between the two values $A_+$ and $A_-$ with transition rates
$r_+$ and $r_-$ \cite{horsthemke, gardiner}. $r_+$ denotes the rate for changing from $A_+$ to
$A_-$ and $r_-$ denotes the rate of passage from $A_-$ to $A_+$.  The mean value of this process,
i.e. $\langle \zeta(t)\rangle =(r_{-}A_++r_{+}A_-)/(r_{+}+r_{-})$, is
fixed at zero in what follows, since the corresponding mean value can
always be incorporated into the term $\Omega$.

At first, we derive an explicit expression for the effective diffusion
coefficient in our system which is defined as the long-time limit
\begin{equation}
  D_{\text{eff}}=\lim_{t\rightarrow\infty}\frac{\langle (\vec{r}(t)-\vec{r}_0)^2\rangle}{4t}\,.
\label{deff}
\end{equation}
We then discuss its dependence on the DMP parameters and on the
external noise for the case without torque as well as for the case
with non-vanishing torque $\Omega$.  We observe a torque-induced
non-monotonic behavior of the effective diffusion coefficient, similar
to the one recently discussed in the context of a system which is
driven by an Ornstein-Uhlenbeck process (OUP) \cite{weberoup}, as well
as to the peaked diffusion of spiral waves driven by a correlated
random forcing \cite{sendina}. Finally, we study the DMP limits
leading to shot noise and to Gaussian white noise \cite{vanBroeck}.
The latter reproduces the well known result which was previously
derived in \cite{schimansky05,teeffelen,weberoup}.

\textbf{Analytical considerations of a DMP-driven agent.}\;\; In Fig.\ref{phiDMNtraj}, we present
the spatial trajectories as obtained from
simulations of our dynamics [Eq.(\ref{curvesys})] for vanishing torque,
small noise intensity $D_{\xi}$ and different parameters of the
dichotomous noise.  The trajectories clearly show the dominant circular
structures, as well as the strong influence of the parameters of the
DMP on the particle's displacement. The effective diffusion coefficient of particles moving at a constant speed $v_0$, can be written with the Taylor-Kubo relation as
\begin{align}
  D_{\text{eff}} =&\;\frac{1}{2}\int_0^{\infty}\langle \vec{v}(t)\vec{v}(t+\tau)\rangle\mathrm d\tau = \frac{v_0^2}{2}\int_0^{\infty}\langle \mathrm{cos}(\Delta (\tau))\rangle\mathrm d\tau\nonumber\\
  =&\;\frac{v_0^2}{2}
  \operatorname{Re}\left(\int_0^{\infty}\int_{-\infty}^{\infty}e^{i\Delta}
    P(\Delta, \tau)\mathrm d\Delta  \mathrm d\tau\right) ,
\label{vgldmnc}
\end{align}
where $P(\Delta,\tau)$ denotes the probability for an angle increment
$\Delta$ during time $\tau$. Using a continuous-time generalization of the classical persistent
random walk, as it was studied in \cite{continuousrand}, we derive the
probability to finish a dichotomous Markov step at a certain time with
a certain angle. This calculation can be explicitly done
\cite{schimansky08} and leads to a coupled system of integral
equations, which can be solved in an algebraic way by considering the
Fourier-Laplace transform
\begin{equation}
 \tilde P(k,s)=\int_{-\infty}^{\infty}e^{ik\Delta} \int_0^{\infty} e^{-s\tau} P(\Delta,\tau)\mathrm d\tau\mathrm d\Delta 
\label{foulap}
\end{equation}
of the corresponding probability density. With $\tilde P(k,s)$,
Eq.(\ref{vgldmnc}) can be rewritten as
\begin{equation}
 D_{\text{eff}}= \frac{v_0^2}{2} \operatorname{Re} \big(\tilde P(k=1,s=0)\big) ,
\end{equation}
and we are finally able to derive the effective diffusion coefficient
of our dynamics
\begin{widetext}
\begin{multline}
  D_{\text{eff}}=\frac{v_0^2}{2} \biggl[
  \frac{(D_{\xi}+r_++r_-)(D_{\xi}(D_{\xi}+r_++r_-)-(A_++\Omega)(A_-+\Omega))}
  {(D_{\xi}(D_{\xi}+r_++r_-)-(A_++\Omega)(A_-+\Omega))^2 +
    ((A_++\Omega)(D_{\xi}+r_-) + (A_-+\Omega)(D_{\xi}+r_+))^2}\\
  + \frac{\left(\frac{A_+r_++A_-r_-}{r_++r_-} +
      \Omega\right)((A_++\Omega)(D_{\xi}+r_-)+(A_-+\Omega)(D_{\xi}+r_+))}{(D_{\xi}(D_{\xi}+r_++r_-)-(A_++\Omega)(A_-+\Omega))^2
    + ((A_++\Omega)(D_{\xi}+r_-) +
    (A_-+\Omega)(D_{\xi}+r_+))^2}\biggl] .
\label{deffDMNcurvetot}
\end{multline}
Without loss of generality we assume that $A_+>0>A_-$ and
$|A_+|>|A_-|$ which also implies $r_+>r_-$. For describing the DMP
dynamics in a more intuitive way, we use the following parameters
\cite{Cebiroglu}
\begin{equation}
 A=A_+-A_- ,  \quad\tau_c=\frac{1}{r_+ + r_-}, \quad 0<p=-\frac{A_-}{A_+}\leq1 .
\label{paraDMNdeff}
\end{equation}
Now, $A$ measures the strength of the process and $\tau_c$ is the
correlation time of the DMP. The parameter $p$ controls the asymmetry
of the driving, the symmetric case corresponds to $p=1$, and it tends
to $0$ for strongly asymmetric strokes.

Rewriting Eq.(\ref{deffDMNcurvetot}) in terms of these new parameters, we get
\begin{equation}
  D_{\text{eff}}=\frac{v_0^2}{2}\cdot \frac{\left(D_\xi + \frac{1}{\tau_c}\right)\left(D_{\xi}^2 + \frac{D_{\xi}}{\tau_c} + \frac{p A^2}{(1+p)^2} - \frac{\Omega A (1-p)}{1+p} - \Omega^2\right) + \left(\frac{1-p}{1+p}A+\Omega\right)\left(\frac{\Omega}{\tau_c} + 2\Omega D_{\xi} +\frac{D_{\xi}A(1-p)}{1+p}\right)}{\left(D_{\xi}^2 + \frac{D_{\xi}}{\tau_c} + \frac{p A^2}{(1+p)^2} - \frac{\Omega A(1-p)}{1+p} - \Omega^2\right)^2 + \left(\frac{\Omega}{\tau_c} + 2\Omega D_{\xi} +\frac{D_{\xi}A(1-p)}{1+p}\right)^2} .
\label{deffparam}
\end{equation}
\end{widetext}

\begin{figure}[t]
\centering
\includegraphics[width=0.455\textwidth]{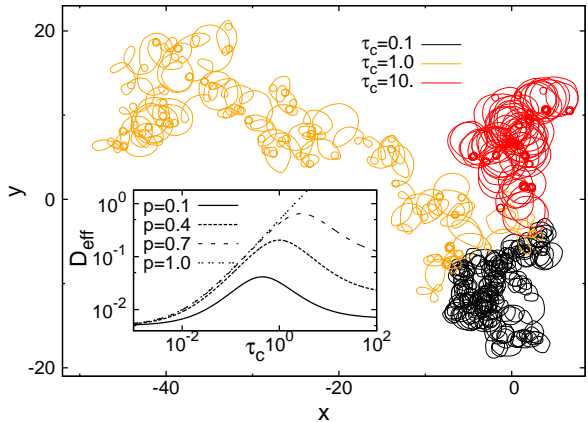}
\caption{\small (Color online) Spatial trajectories for different
  $\tau_c$ values and a non-vanishing torque $\Omega=1$; the other
  parameters are $p=0.4$, $D_{\xi}=0.01$, $A=2$, $v_0=1$, and total duration
  $t_l=10$; the inset illustrates the effective diffusion coefficient
  $D_{\text{eff}}$ versus correlation time $\tau_c$ for different
  asymmetry parameters $p$. }
\label{trajtau}
\end{figure}
\begin{figure}
\centering
\includegraphics[width=0.5\textwidth]{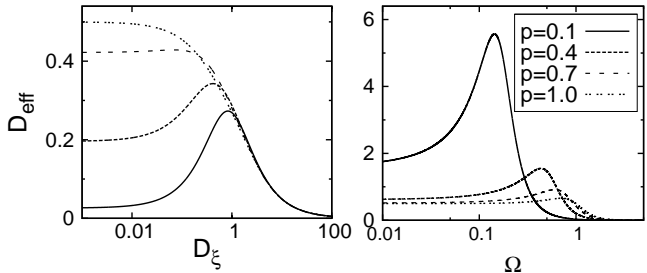}
\caption{\small Effective diffusion coefficient versus noise intensity
  $D_{\xi}$ (left) and torque $\Omega$ (right) for different asymmetry
  parameters $p$ within the theory (cf. Eq.(\ref{deffparam}))
  for the constant torque $\Omega=1$(left), $D_{\xi}=0.01$(right),
  $\tau_c=1$, $v_0=1$, $A=2$.}
\label{torqueDMN1}
\end{figure}

\textbf{Discussion of the effective diffusion coefficient.}\;\;
Considering a vanishing torque $\Omega=0$, we can discuss the
symmetric and asymmetric limit of our DMP drive. A strong asymmetry
leads with Eq.(\ref{deffparam}) to
\begin{equation}
 \lim_{p\rightarrow 0} D_{\text{eff}}=\frac{v_0^2}{2D_{\xi}} ,
\label{Deffdmncurve}
\end{equation}
so that we receive a divergent expression for a vanishing intensity of
the thermal noise $\xi(t)$. Hence, the noise leads to the maintenance
of the diffusive character of our system. This limit coincides
with the result of the Gaussian white noise-driven angle discussed in \cite{mikhailov}. In the limit
$p\rightarrow 0$, the DMP-term
$\zeta(t)$ becomes negligible for the simple reason that the torque $A_-$ vanishes. Therefore, only the Gaussian force $\xi(t)$ drives the angle which results in Eq.(\ref{Deffdmncurve}).

The symmetric, torqueless case $p=1$ on the other hand, reproduces the
result
\begin{equation}
 \lim_{p\rightarrow 1} D_{\text{eff}}=\frac{v_0^2}{2}\cdot \frac{D_{\xi}+\frac{1}{\tau_c}}{D_{\xi}^2 + \frac{D_{\xi}}{\tau_c} + \frac{A^2}{4}} ,
\label{symmlim}
\end{equation}
which has been previously derived in \cite{schimansky08}. Analyzing
this expression leads to a noise intensity which maximizes the
effective diffusion coefficient, namely the value which obeys
$D_{\xi}+1/\tau_c=A/2$. The two dissipative time scales of the
noises sources, i.e. $\tau_{\xi}= 1/D_{\xi}$ and $\tau_c$, relate to the
torque of the DMP which turns the particle.

A non-vanishing torque $\Omega\neq 0$ changes the basic
characteristics of our dynamics and induces a systematic change of the curvature of the path. Figure \ref{trajtau} shows
spatial trajectories for different values of $\tau_c$ and illustrates
the influence of $\Omega$. The fact that this influence on $D_{\text{eff}}$ leads to a
non-monotonous parameter dependence is demonstrated in Fig.\ref{trajtau}(inset) and
\ref{torqueDMN1}.  We recognize that $D_{\text{eff}}$ as a function of
$\tau_c$, $D_{\xi}$ and $\Omega$ possesses a well-defined maximum (a
similar $p$ dependence is not shown here).

This can be understood in view of the corresponding behavior of
the agent considered [see Fig.\ref{trajtau}].  Small and large
correlation times lead to a curled structure where the system stays in
a certain DMP state either too long or too short in order to perform a
considerable spatial displacement. In the case of $\tau_c=10$, the
particle performs a persistent circular motion with small curvature
$(\Omega+A_-)$ interrupted by small spins with large curvature
$(\Omega+A_+)$ and the additive noise causes diffusion by shifting
the centers of the circles stochastically. For the case with
$\tau_c=0.1$, fast DMP-switches induce an erratic motion. In
contrast to the torqueless situation, the non-vanishing $\Omega$ reduces the
displacement of the particle. On average, the motion follows again
randomized circular lines determined by the non-vanishing torque and
the fast DMP strokes, whose mean influence disappears for fast
switchings. Calculating the limits of large and small $\tau_c$ values analytically, results in
a non-zero value of $D_{\text{eff}}$ in both cases due to the
additive Gaussian noise.  This property is seen in Fig.\ref{trajtau} where both
asymptotics tend to finite values.

The optimal $\tau_c$ in between the two limits, i.e. the state of maximal diffusion, corresponds to
a maximally stretched trajectory for given values $A_{\pm}$ and only
the $A_-$ stroke can decrease the curvature. Thereby, it can
induce longer excursions which become maximal if the mean waiting time $1/r_-$ matches the
time which the angle $\phi$ needs to rotate over half of a circle
during the $A_-$ stroke. Hence, the parameters have to obey
$|A_-+\Omega|/r_-= \pi$ and for the notation introduced in Eq.
(\ref{paraDMNdeff}) follows
\begin{equation}
  \tau_{c}^{\text{max}}=\frac{\pi p}{|\Omega(1+p)-pA|}.
\label{tauapp}
\end{equation}
This result is in good agreement with the peaks in Fig.\ref{trajtau}
for small $p$ values. It fails for larger $p$, where the influence of
the $A_+$ stroke is not negligible.

The behavior of $D_{\text{eff}}$ as a function of $\Omega$
[Fig.\ref{torqueDMN1}(right)] shows a peak due to similar reasons. The
case $\Omega=|A_-|$ causes straight paths within the corresponding DMP
mode and will therefore enhance the spread. The peak in the dependence of $D_{\text{eff}}$
on the noise intensity $D_{\xi}$ in Fig.\ref{torqueDMN1}(left) shifts
for growing $p$ to smaller noise values. Such a peak was already
reported for similar dynamics but in the absence of a DMP \cite{schimansky05,teeffelen}. It is in
agreement with our previous discussion of Eq.(\ref{Deffdmncurve})
that the effect of the DMP strokes disappear if $p \to 0$ .

\begin{figure}
\centering
\includegraphics[width=0.455\textwidth]{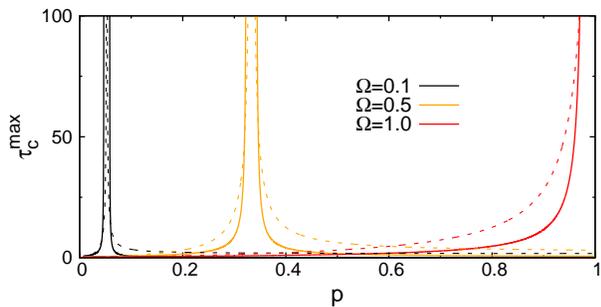}
\caption{\small (Color online) $\tau_c$ value which maximizes the effective diffusion coefficient versus asymmetry parameter $p$ for different mean torques $\Omega$; the results of the full theory in Eq.(\ref{deffparam}) (solid lines) and for the approximative expression in Eq.(\ref{tauapp}) (dashed lines) are shown for the noise intensity $D_{\xi}=0.01$ and the stroke strength $A=2$.}
\label{maxDMN}
\end{figure}

Taking the derivative of Eq.(\ref{deffparam}) with respect to $\tau_c$ leads, after some
straightforward calculations, to a lengthy analytical result. $\tau_c$ values which maximize the diffusion coefficient are presented in Fig.\ref{maxDMN}. It shows
a perfect agreement with simulation results and with the peaks in
Fig.\ref{trajtau}.  The rough approximation given by Eq.(\ref{tauapp})
turns out to be rather good.
The peaks in Fig.\ref{maxDMN} occur because of the mentioned rectilinear
motion for $\Omega=|A_-|=pA/(1+p)$, since an increase of $\tau_c$ also
increases the duration of the rectilinear motion and enlarges
consequently $D_{\text{eff}}$. That is why the peak for $p=1$ is
shifted to infinitely large correlation times in Fig.\ref{trajtau}(inset).

The white shot noise limit of the DMP drive can be found by considering the
limits $A_+\rightarrow\infty$ and $r_+\rightarrow\infty$ while the
ratio $A_+/r_+=-A_-/r_-=w$ holds constant \cite{vanBroeck}. The
corresponding locomotion of the agent consists of a circular motion
with mean curvature $(\Omega+A_-)$ interrupted by infinitely fast
turnings of the angle, induced by the short, large strokes $A_+$. The
autocorrelation function of $\zeta(t)$ in this shot noise limit reads $\langle \zeta(t)\zeta(t+\tau)\rangle=(-A_-w/\tau_c)\, \mathrm{exp}\left(-|\tau|/\tau_c\right)$ and therefore
implies the noise intensity $D_{\zeta}=-A_-w$, while $A_-<0$ holds. Rewriting of
Eq.(\ref{deffDMNcurvetot}) within the mentioned limits leads to the effective diffusion coefficient
\begin{equation}
D_{\text{eff}}^{\text{shot}}=\frac{v_0^2}{2}\cdot \frac{D_{\xi} + D_{\zeta} + D_{\xi}w^2}{(D_{\xi} + D_{\zeta} - \Omega w)^2 + (D_{\xi}w + \Omega)^2} .
\label{shotexp}
\end{equation} 
Here, $D_{\text{eff}}^{\text{shot}}$ becomes maximal for
$\Omega=-A_-w^2/(1+w^2)$, which implies $\Omega \le
|A_-|$.

The white Gaussian limit of the DMP drive can be derived by considering the limit $w\rightarrow 0$ while $D_{\zeta}$ is hold constant \cite{vanBroeck}. Doing
so in Eq.(\ref{shotexp}), we find
\begin{equation}
 D_{\text{eff}}^{\text{Gauss}}=\frac{v_0^2}{2}\cdot \frac{D_{\xi} + D_{\zeta}}{(D_{\xi} + D_{\zeta})^2 + \Omega^2}.
\end{equation}
If we introduce a total noise intensity $D=D_{\xi} + D_{\zeta}$, this
expression coincides with the result in
\cite{schimansky05,teeffelen,weberoup} for an agent under influence of Gaussian
white noise with intensity $D$. Maximal diffusion at the value
$D_{\text{eff}}^{\text{Gauss}}=v_0^2/(4\Omega)$ is obtained for the
total noise intensity $D=\Omega$.  Thus, the
resonance occurs where $D$, i.e. the angular correlation decay rate in
the case of a Gaussian white angle drive, equals the effective torque,
as it is likewise the case in Eq.(\ref{symmlim}) with the additional
decay rate of the correlation within the DMP drive.

\textbf{Conclusion.}\;\; We have discussed exact results for the effective
diffusion coefficient of a particle moving at a constant speed under
influence of a constant torque, dichotomous angular Markov noise and
additional directional Gaussian perturbations. The results help to
understand the behavior of our system in a qualitative and
quantitative way. They clarify the role of the asymmetry and of an
additional torque in the DMP-driven angle dynamics.  The strongly
peaked bimodal angular probability distribution, which we have assumed
in our model, is of course an enormous simplification of the ones
found in real biological systems.  But in view of the bulk of works
discussing symmetric angular distributions
\cite{peruani,mikhailov,schimansky08,othmer,okubo}, it seems to be
reasonable to discuss the influence of a certain asymmetry as
well. Since the diffusion coefficient is one of the most easily accessible
quantities in experiments, we hope that this work not only fills a gap in our general
theoretical understanding of self-propelled agents, but will also
stimulate corresponding experimental studies.

\textbf{Acknowledgments.}\;\; This work has been supported by the DFG
- IRTG 1740. C.W. acknowledges a scholarship from the German National Academic Foundation. We also thank Patrick Lessmann for fruitful discussions.


\begin{thebibliography}{99}
\bibitem{daphniapattern} A. Ordemann and G. Balazsi and F. Moss, Physica A
  \textbf{325}, 260 (2003).
\bibitem{schimansky08} L. Haeggqwist, L. Schimansky-Geier, I.M. Sokolov, and
  F. Moss, Eur. Phys. J.-- ST {\bf 157}, 33 (2008).
\bibitem{schimansky04} N. Komin, U. Erdmann, and L. Schimansky-Geier, Fluctuation and Noise Letters \textbf{4}, 151 (2004).
\bibitem{pavelswarm} P. Romanczuk and I.D. Couzin and L.
  Schimansky-Geier, Phys. Rev. Lett. \textbf{102},
  10602 (2009).
\bibitem{friedrich2} B.M. Friedrich and F. J{\"u}licher, New J.
 Phys. \textbf{10},  123025 (2008).
\bibitem{kanokov} Z. Kanokov, J.W.P.  Schmelzer,  and A.K. Nasirov,
  CEJP \textbf{8}, 667 (2010).
\bibitem{gautrais} J. Gautrais, C. Jost, M. Soria, A. Campo, S. Motsch, R.
  Fournier, S. Blanco, and G. Theraulaz, J.  Math.  Biol.
  \textbf{58}, 429 (2009).
\bibitem{horsthemke} W. Horsthemke and R. Lefever, {\it Noise
    Induced Transitions, Theory and Applications in Physics, Chemistry
    and Biology} (Springer, Berlin, 1983).
\bibitem{gardiner} C. Gardiner, \textit{Handbook of stochastic methods} (Springer Berlin, 1985).
\bibitem{weberoup} C. Weber, P.K. Radtke, L. Schimansky-Geier, and P. H\"anggi, Phys. Rev. E \textbf{84} 011132 (2011).
\bibitem{sendina} I. Sendi{\~n}a-Nadal, S. Alonso, V.
  P{\'e}rez-Mu{\~n}uzuri, M.  G{\'o}mez-Gesteira, V. P{\'e}rez-Villar,
  L. Ram{\'i}rez-Piscina, J.  Casademunt, J. M. Sancho, and F.
  Sagu{\'e}s, Phys. Rev. Lett. \textbf{84}, 2734 (2000).
\bibitem{vanBroeck} C. Van Den Broeck, J. Stat. Phys. \textbf{31}, 467 (1983).
\bibitem{schimansky05} L. Schimansky-Geier, U. Erdmann, and N. Komin,
  Physica A \textbf{351}, 51 (2005). 
\bibitem{teeffelen} S. van Teeffelen and H. L\"owen, Phys. Rev. E {\bf
    78}, 020101 (2008).
\bibitem{continuousrand} J. Masoliver, K. Lindenberg, and G. Weiss, Physica A \textbf{157}, 891 (1989).
\bibitem{Cebiroglu} G. Cebiroglu, C. Weber, and L. Schimansky-Geier, Chem. Phys. \textbf{375}, 439 (2010).
\bibitem{mikhailov} A.S. Mikhailov and D. Meink{\"o}hn, in {\em
    Stochastic Dynamics}, edited by L. Schimansky-Geier and T. P\"oschel,
  p.  334 (Springer, Berlin, 1997).
\bibitem{othmer} H.G. Othmer, S.R.  Dunbar and W. Alt, J.
  Math. Biology \textbf{26}, 263 (1988).
\bibitem{okubo} A. {\=O}kubo and S.A. Levin, \textit{Diffusion and ecological problems: modern perspectives} (Springer, 2002).
\bibitem{peruani} F. Peruani and L.G. Morelli, Phys. Rev. Lett.
  \textbf{99}, 10602 (2007).

\end{thebibliography}
\end{document}